\documentstyle[11pt]{article}
\def\g{{\,\rm \gamma \,}}

\def\Id{{\,\rm Id \,}}

\def\CC{{\,\rm C\,}}

\def\o{{\,\rm o \,}}

\def\sym{{\,\rm sym \,}}

\def\ii{{\,\rm i \,}}
\def\dist{{\,\rm dist \,}}

\def\SO{{\,\rm SO \,}}

\def\B{{\,\cal B \,}}

\def\+M{{\,\rm M^{n\times n}_+ \,}}
\def\tr{{\,\rm tr \,}}

\def\qfq{{\quad\mbox{for}\quad}}

\def\ii{{\,\rm i \,}}

\def\lam{\lambda}

\def\E{{\cal E}}

\def\E{{\cal E}}

\newfont{\Blackboard}{msbm10 scaled 1200}

\newfont{\roma}{cmr10 scaled 1200}

\def\<{{\langle}}
\def\>{{\rangle}}

\def\g{\gamma}
\def\Ga{\Gamma}
\def\var{\varphi}
\def\si{\sigma}

\def\a{\alpha}
\def\b{\beta}

\def\Om{\Omega}

\newtheorem{thm}{{}\hskip\parindent Theorem}[section]
\newtheorem{lem}{{}\hskip\parindent Lemma}[section]
\newtheorem{pro}{{}\hskip\parindent Proposition}[section]

\newtheorem{rem}{{}\hskip\parindent Remark}[section]

\def\dfrac{\displaystyle\frac}

\def\pl{\partial}
\def\rw{\rightarrow}

\def\be{\begin{equation}}
\def\ee{\end{equation}}
\def\beq{\arraycolsep=1.5pt\begin{eqnarray}}
\def\eeq{\end{eqnarray}}

\def\R{I\!\!R}

\def\n{\vec{n}}
\large
\title{Lower Bounds of Optimal Exponentials of Thickness in  Geometry Rigidity Inequality for Shells}
\date{}
\author{
Peng-Fei YAO\\[0.3cm]
Key Laboratory of  Systems and Control\\
Institute of Systems Science,
Academy of Mathematics and Systems Science\\
Chinese Academy of Sciences, Beijing 100190, P. R. China\\
School of Mathematical Sciences\\
University of Chinese Academy of Sciences, Beijing 100049, China\\
e-mail: pfyao@iss.ac.cn}

\begin{document}
\maketitle
 \footnote{This work is  supported by the National
Science Foundation of China, grants  no. 61473126 and no. 61573342, and Key Research Program of Frontier Sciences, CAS, no. QYZDJ-SSW-SYS011.}

\begin{quote}
\begin{small}
{\bf Abstract} \,\,\,The optimal exponentials of the thickness in the geometry rigidity inequality of shells represent the geometry rigidity of the shells. We obtain that the lower bounds of the optimal exponentials are $4/3,$ $3/2,$ and $1,$ for the hyperbolic shell, the parabolic shell, and the elliptic shell, respectively, through  the construction of  the  Ans\"{a}tze.
\\[3mm]
\end{small}
\end{quote}

\setcounter{equation}{0}
\section{Introduction and Main Results}
\def\theequation{1.\arabic{equation}}
\hskip\parindent The geometry rigidity inequality,  namely the Friesecke-James-Muller  estimate \cite{FrJaMu,FrJaMu1},
plays a central role in models in nonlinear elasticity. In their basic form, these estimates assert
that for a deformation $u\in H^1(\Om,\R^n)$ the distance of $\nabla u$ to a suitably chosen proper rotation
$Q\in\SO(n)$ is dominated in $L^2$ by the distance function of $\nabla u$ to $\SO(n).$ The proof \cite{FrJaMu} is
based on the fact that the nonlinear estimate can be related to the linear one since the tangent
space to the smooth manifold $\SO(n)$ at the identity matrix is given by the linear space of all
skew-symmetric matrices.
In fact, geometric rigidity results are the cornerstone of rigorous derivations of two dimensional
plate and shell theories from three-dimensional models in the framework of
nonlinear elasticity theory. The $L^2$ version by Friesecke et al. \cite{FrJaMu} generalized previous work \cite{John, John1, Kohn,Rese, Rese1}
and allowed for the first time the derivation of limiting theories as the thickness
of the three-dimensional structure tends to zero without a priori assumptions on the
deformations in various scaling regimes \cite{FJMM,FrJaMu,FrJaMu1,HoLePa,LeMoPa,LePa,LeMoPa1,Yao2017} and many others.

It is known that the rigidity of a shell  is closely related
to the optimal constant of thickness in the geometric rigidity estimate \cite{CM,DS,FrJaMu1,GH2} and the optimal constant is crucial to  shell theories being derived from
$3$-dimensional elasticity by $\Ga$-convergence like \cite{FJMM,FrJaMu,HoLePa,LeMoPa,LePa,LeMoPa1,Yao2017}. As a linear version of the geometric rigidity estimate,  the the optimal constants of thickness in  Korn's inequalities have been calculated subject to the Gaussian curvature of the middle surface of a shell under the assumption that the middle surface is given by a single principal curvature coordinate \cite{GH,GH1,Ha2}. This assumption that the middle surface being a single principal curvature coordinate is generalized in \cite{Yao20181,Yao2018}. In  Korn's inequalities the optimal constant for the plate calculated in \cite{FrJaMu1} scales like $h^2,$ for  cylindrical shells in \cite{GH}, $h^{3/2},$ for the positive curvature in \cite{Ha2},
$h,$ and for the negative curvature in \cite{Ha2}, $h^{4/3},$ respectively. It is expected that the
analogous nonlinear estimates  will have the same scaling of the constant  in terms of the shell thickness.

Here we calculate some lower bounds of the  optimal exponentials of the  thickness in  the nonlinear geometry rigidity inequality subject to the curvature of the middle surface.

Let $M\subset\R^3$ be a $\CC^3$ surface with the induce metric $g$ and a normal field $\n.$ Let $S\subset M$ be an open, simply connected, bounded set with a regular boundary $\pl S.$ We consider a shell with thickness $h>0$
$$\Om=\{\,p+t\n(p)\,|\,p\in S,\,-h/2<t<h/2\,\}.$$
Let $\kappa$ be the Gaussian curvature of $M.$ We say that $\Om$ is parabolic if
\be\kappa(p)=0,\quad|\Pi(p)|>0\qfq p\in\overline{S},\label{1.2}\ee where $\Pi=\nabla\n$ is the second fundamental form of $M.$
If
\be\kappa(p)>0\qfq p\in\overline{S},\label{1.6}\ee
then $\Om$ is said to be elliptic. In addition, $\Om$ is said to be hyperbolic if
$$\kappa(p)<0\qfq p\in\overline{S}.$$

For $A\in\R^{3\times3},$ we denote the Euclidean norm by $|A|=\sqrt{{\tr AA^T}}.$ The distance
from $A$ to $\SO(3)$ is denoted $\dist(A,\SO(3)).$
Let $\mu>0$ be such that  estimate (\ref{1.1}) below holds true. There is a constant $C>0,$ independent of $h>0,$ such that for every
$u\in H^1(\Om,\R^3)$ there exists a constant rotation $Q\in\SO(3),$ such that
\be\|\nabla u-Q\|^2_{L^2(\Om)}\leq\frac{C}{h^\mu}\int_\Om\dist^2(\nabla u(z),\SO(3))dz.\label{1.1}\ee
Set
$$\mu(\Om)=\inf\{\,\mu\,|\,\mu>0\,\mbox{such that (\ref{1.1}) holds}\,\}.$$
It follows from \cite{FrJaMu1} that
$$\mu(\Om)=2$$ if $\Om$ is a plate, and from \cite{LeMoPa} that
$$\mu(\Om)\leq2$$ for a shell, respectively.

We have the following.

\begin{thm}\label{t1.1}
If $\Om$ is hyperbolic, then
\be\mu(\Om)\geq4/3.\label{e2}\ee
In the case of $\Om$ being elliptic,
\be\mu(\Om)\geq1.\label{e1}\ee
\end{thm}

Next, we consider the case of the parabolic. We need

\begin{pro}\label{p1}Let $M\subset\R^3$ be a parabolic surface without boundary. Then for given $p\in M,$
there exists a unique regular geodesic $\g(t,p)$ on $M$ such that
\be\g(0,p)=p,\quad |\dot\g(t,p)|=1,\quad \nabla_{\dot\g(t,p)}\n=0\qfq t\in\R.\label{pra}\ee Moreover, $\g(t,p)$ is a straight line in $\R^3.$
\end{pro}

\begin{thm}\label{t1.2} Let $M\subset\R^3$ be a parabolic surface and let $S\subset M$ be a bounded open set. We suppose that there is a point $p_0\in S$ such that the following assumption holds: Let $t_-<0<t_+$ be such that
\be\g(t_\pm,p_0)\in\pl S,\quad(\nabla_{\tau_\pm}\n)(\g(t_\pm,p_0))\not=0,\quad \g(t,p_0)\in S \qfq t\in(t_-,t_+),\label{ga7}\ee where $\g$ is given in $(\ref{pra})$ and
$\tau_\pm\in T_{\g(t_\pm,p_0)}\pl S$ with $|\tau_\pm|=1.$ Then
\be\mu(\Om)\geq3/2.\label{e3}\ee
\end{thm}

\begin{rem}
It is conjectured that all the equal signs in $(\ref{e2}),$ $(\ref{e1}),$ and $(\ref{e3})$ hold.

\end{rem}

The estimates (\ref{e2}), (\ref{e1}), and (\ref{e3}) will be obtain by constructing the Ans\"{a}tze. The main observation is that such Ans\"{a}tzes  may come from the improvement in the ones for the Korn inequality. In the case of the Korn inequality the Ans\"{a}tzes are constructed in \cite{GH}, \cite{GH1},  and \cite{Ha2}, which are based on the main assumption that the middle surface is given by a single principal curvature coordinate, i.e.,
\be S=\{\,{\bf r}(z,\theta)\,|\,(z,\theta)\in[1,1+l]\times[0,\theta_0]\,\},\label{as}\ee where the properties
$$\nabla_{\pl z}\n=\kappa_z\pl z,\quad\nabla_{\pl\theta}\n=\kappa_\theta\pl\theta\qfq p\in S $$ hold. In the case of the parabolic or hyperbolic shell, a principal coordinate only exists locally (\cite{Yao2018}). There is even no such a local existence for the elliptic shell.

Here we will construct  the Ans\"{a}tze for the korn inequality  without assumption (\ref{as}) and then improve them to obtain the ones for the geometric rigidity estimate.

\setcounter{equation}{0}
\section{Proof of the Main Results}
\def\theequation{2.\arabic{equation}}
\hskip\parindent Let $\nabla$ and $D$ denote the connection of $\R^3$ in the Euclidean metric and the one of $M$ in the reduced metric, respectively. We have to treat the relationship between $\nabla$ and $D$ carefully.

We need a linear operator $Q$ as follows. Let $M$ be oriented and $\E$ be the volume element of $M$ with the positive orientation. Let $p\in M$ be given and let $e_1,$ $e_2$ be an orthonormal basis of $T_pM$  with positive orientation, that is,
$$\det\Big(e_1,e_2,\n(p)\Big)=1.$$ We define $Q:$ $T_pM\rw T_pM$ by
\be Q\a=\<\a,e_2\>e_1-\<\a,e_1\>e_2\quad\mbox{for all}\quad\a\in T_pM.\label{1.3n}\ee $Q$ is well defined in the following sense: Let $\hat e_1,$ $\hat e_2$ be a different orthonormal basis of $T_pM$  with  positive orientation,
$$\det\Big(\hat e_1,\hat e_2,\n(p)\Big)=1.$$ Let
$$\hat e_i=\sum_{j=1}^2\a_{ij}e_j\qfq i=1,\,\,2.$$ Then
$$1=\E(\hat e_1,\hat e_2)=\a_{11}\a_{22}-\a_{12}\a_{21}.$$ Using the above formula, a simple computation yields
$$ \<\a,\hat e_2\>\hat e_1-\<\a,\hat e_1\>\hat e_2=\<\a, e_2\> e_1-\<\a, e_1\> e_2.$$ Clearly, $Q:$ $T_pM\rw T_pM$ is an isometry and
$$Q^T=-Q,\quad Q^2=-\Id.$$
The operator $Q$ plays an important role in the case of the hyperbolic surface \cite{Yao2017}.

Let $S$ be hyperbolic and $p_0\in S$ be given. Let $\psi(p)=(x_1,x_2):$ $\B(p_0,3\delta)\rw\R^2$ be an {\it asymptotic coordinate system} with $\psi(p_0)=0$
\be\Pi(\pl x_1,\pl x_1)=\Pi(\pl x_2,\pl x_2)=0\qfq p\in \B(p_0,3\delta),\label{o}\ee where $\B(p_0,3\delta)\subset S$ is the geodesic ball centered at $p_0$ with radius $3\delta$ where $\delta>0$ is small. It is further assumed that $\psi(p)=x$ is positively orientated, i.e.,
\be\det\Big(\pl x_1,\pl x_2,\n(p)\Big)>0\qfq p\in\B(p_0,3\delta).\label{po}\ee
Let $\var\in\CC^1_0(\B(p_0,3\delta))$ be given such that
$$\var(p)=1\qfq p\in \B(p_0,2\delta).$$
We define
\be f(p)=\var(p)x_1(p)\qfq p\in S,\label{f}\ee where $x_1$ is the first component of the $\psi(p)=(x_1,x_2)$ for $p\in \B(p_0,3\delta).$ Then
$$f(p)=0\qfq p\in S/\B(p_0,3\delta).$$

\begin{lem}\label{lQ1} Let $S$ be hyperbolic and $f$ be given in $(\ref{f}).$  Then
\be \Pi(QDf,QDf)(p)=0\qfq p\in\B(p_0,2\delta),\label{Q1}\ee where $D$ is the connection of $S$ in the induced metric $g.$
\end{lem}

{\bf Proof}\,\,\,Consider the asymptotic coordinate system (\ref{o}). Set
\be E_1=a_1\pl x_1,\quad E_2=b_1\pl x_1+b_2\pl x_2,\label{Q}\ee where
$$a_1=\frac{1}{\sqrt{g_{11}}},\quad b_1=-\frac{g_{12}}{\sqrt{g_{11}\det G}},\quad b_2=\sqrt{\frac{g_{11}}{\det G}},$$ where
$$G=\Big(g_{ij}\Big),\quad g_{ij}=\<\pl x_i,\pl x_j\>.$$
Then $E_1,$ $E_2$ forms an orthonormal frame with positive orientation on $\B(p_0,\delta)$ since it follows from (\ref{po}) that
$$\det\Big(E_1,E_2,\n(p)\Big)=\det\Big(\pl x_1,\pl x_2,\n(p)\Big)\left(\begin{array}{ccc}a_1&b_1&0\\
0&b_2&0\\
0&0&1\end{array}\right)=a_1b_2\det\Big(\pl x_1,\pl x_2,\n(p)\Big)=1$$ for $p\in\B(p_0,\delta).$ Using (\ref{1.3n}) and (\ref{Q}), we have
$$\left\{\begin{array}{l}Q\pl x_1=-\dfrac1{a_1}(b_1\pl x_1+b_2\pl x_2),\\
Q\pl x_2=\dfrac1{b_2}(a_1+\dfrac{b_1^2}{a_1})\pl x_1+\dfrac{b_1}{a_1}\pl x_2.\end{array}\right.$$

From (\ref{f}),
$$f_{x_1}=1,\quad f_{x_2}=0\qfq p\in \B(p_0,2\delta).$$ We thus obtain
\be Df=g^{11}\pl x_1+g^{12}\pl x_2,\label{pl}\ee where $\Big(g^{ij}\Big)=\Big(g_{ij}\Big)^{-1},$ and
\beq(\det G)QDf&&=\det G[\frac{g^{12}}{a_1b_2}(a_1^2+b_1^2)-\frac{g^{11}b_1}{a_1}]\pl x_1+\frac{\det G}{a_1}(g^{12}b_1-g^{11}b_2)\pl x_2\nonumber\\
&&=-\sqrt{\det G}\pl x_2\qfq p\in\B(p_0,2\delta).\label{Q2}\eeq

(\ref{Q1}) follows from (\ref{o}) and (\ref{Q2}). \hfill$\Box$

\begin{lem}\label{lQ2}Let $S$ be hyperbolic. Then the shape operator $\nabla\n:$ $T_pS\rw T_pS$ is reversible. Let $\kappa$ be the Gaussian curvature and let $f$ be given in $(\ref{f}).$ Set
\be  Z=(\nabla\n)^{-1}[(\nabla\n)^2]^{1/2}Df,\quad v=\frac{|Df|^2}{\sqrt{|\Pi|^2-2\kappa}}\qfq p\in S. \label{Z}\ee Then
\be\sym Z\otimes Df=v\Pi\qfq p\in\B(p_0,2\delta).\label{Pi5}\ee
\end{lem}

{\bf Proof}\,\,\,Let $p\in \B(p_0,\delta)$ be given. Let $e_1,$ $e_2$ be an orthonormal basis of  $T_pS$ with positive orientation such that
\be\nabla_{e_1}\n=\lam_1e_2,\quad\nabla_{e_2}\n=\lam_2e_2,\quad \lam_1>0>\lam_2.\label{Pi2}\ee Then
\be Z=e_1(f)e_1-e_2(f)e_2.\label{Pi3}\ee
Moreover, it follows from Lemma \ref{lQ1} and (\ref{Pi2}) that
\be 0=\Pi\Big(e_2(f)e_1-e_1(f)e_2,e_2(f)e_1-e_1(f)e_2\Big)=\lam_1[e_2(f)]^2+\lam_2[e_1(f)]^2.\label{Pi1}\ee

Using (\ref{Pi2}) and (\ref{Pi1}), we have
\be v=\frac{[e_1(f)]^2+[e_2(f)]^2}{\lam_1-\lam_2}=\frac{[e_1(f)]^2}{\lam_1}=-\frac{[e_2(f)]^2}{\lam_2}.
\label{Pi4}\ee It follows from (\ref{Pi2})-(\ref{Pi4}) that
$$Z\otimes Df(e_1,e_1)=[e_1(f)]^2=v\lam_1=v\Pi(e_1,e_1), $$
$$ Z\otimes Df(e_2,e_2)=-[e_2(f)]^2=v\Pi(e_2,e_2),$$
$$Z\otimes Df(e_1,e_2)=e_1(f)e_2(f),\quad Z\otimes Df(e_2,e_1)=-e_2(f)e_1(f).$$
Thus (\ref{Pi5}) follows from the above formulas. \hfill$\Box$\\

Let $(M,g)$ be a Riemanniann manifold.  Let $T$ be a 2-order tensor field on $(M,g)$ and let $X$ be a vector field on $(M,g).$ We define the inner multiplication of $T$ with $X$ to be another vector field, denoted by $\ii(X)T,$ given by
$$\<\ii(X)T,Y\>=T(X,Y)\qfq Y\in T_pM,\quad p\in M,\quad g=\<\cdot,\cdot\>.$$

For any $y\in H^1(\Om,\R^3),$ we decompose $y$ into
\be y(z)=W(p,t)+w(p,t)\n(p)\qfq z=p+t\n(p)\in\Om,\quad p\in S,\quad |t|<h/2,\label{2.1}\ee where $w=\<y,\n\>$ and $W(\cdot,t)$ is a vector field on $S$ for $|t|<h/2.$ It follows from (\ref{2.1})
that
\be \nabla_{\a+t\nabla\n\a} y=D_\a W+w\nabla_\a\n+[\a(w)-\Pi(W,\a)]\n\qfq \a\in T_pS,\ee
\be\nabla_{\n}y=W_t(p,t)+w_t(p,t)\n(p)\qfq  p\in S,\quad|t|<h/2,\label{2.2}\ee
where $\nabla$ and $D$ are the covariant differentials of the dot metric in $\R^3$ and of the induced metric in $S,$ respectively, and
$W_t=\pl_tW$ and $w_t=\pl_tw.$

By defining $\nabla\n\n=0,$ we introduce an 2-order tensor $P(y)$ on $\R^3_p$ by
\be P(y)(\tilde\a,\tilde\b)=\<\nabla_{\nabla\n\tilde\a}y,\tilde\b\>\qfq\tilde\a,\,\,\tilde\b\in\R^3.\label{2.4}\ee

We have
\begin{lem}(\cite{Yao20181})\label{lem2.1} Let $y\in H^1(\Om,\R^3)$ be given in $(\ref{2.1}).$ Then
\be|\nabla y+tP(y)|^2=|DW+w\Pi|^2+|Dw-\ii(W)\Pi|^2+|W_t|^2+w_t^2,\label{2.5}\ee
\be|\sym\nabla y+t\sym P(y)|^2=|\Upsilon(y)|^2+\frac12|X(y)|^2+w_t^2,\label{2.6}\ee where
\be\Upsilon(y)=\sym DW+w\Pi,\quad X(y)=Dw-\ii(W)\Pi+W_t.\label{2.7}\ee Moreover, the following estimates hold
$$\si|\nabla y|^2\leq|\nabla y+tP(y)|^2\leq C|\nabla y|^2,$$
$$\si|\sym\nabla y|^2-Ch^2|\nabla y|^2\leq|\sym\nabla y+t\sym P(y)|^2\leq C(|\sym\nabla y|^2+h^2|\nabla y|^2),$$ for $h>0$ small, where $|t|\leq h/2.$
\end{lem}

For $A,$ $B\in\R^{3\times3},$ let
\be T(A)=(A^TA)^{1/2}-I,\quad\Phi(B)=\sym B+\frac12B^TB,\label{Phi}\ee where $I$ is the $3\times3$ unit matrix. Then
\be |T(A)|=\dist(A,\SO(3))\qfq A\in\R^{3\times3}\quad\mbox{with}\quad \det A>0.\label{A2}\ee

\begin{lem}\label{l2.4}
\be \frac{|T(A)|}{2\sqrt{3}}\leq|\Phi(B)|\leq\frac{\sqrt{3}+|A|}{2}|T(A)|\qfq\det A>0,\label{2.2}\ee where $A=I+B$ for $B\in\R^{3\times3}.$
\end{lem}

{\bf Proof}\,\,\,(\ref{2.2}) follows from the identities
$$T(A)[(A^TA)^{1/2}+I]=B+B^T+B^TB=2\Phi(B),\quad T(A)=2\Phi(B)[(A^TA)^{1/2}+I]^{-1}.$$\hfill$\Box$

{\bf Proof of  Theorem \ref{t1.1}}\,\,\,(a)\,\,\,Let the middle surface $S$ be hyperbolic. Let $\delta>0$ be given in (\ref{o}). Let $\hat\var\in\CC_0^2(\B(p_0,3\delta))$ be such that
$$\hat\var(p)=1\qfq p\in\B(p_0,\delta),\quad \hat\var(p)=0\qfq S/\B(p_0,2\delta).$$ Let $f$ and $Z,$  $v$ be given in (\ref{f}) and (\ref{Z}), respectively. First, we  look for the ansatze for the Korn inequality in the form
\be y(z)=W(z)+w(p)\n(p),\label{u}\ee where
\be W(z)=\frac{\hat\var(p)\cos(\phi f(p))}{v(p)}Z(p)-tDw(p),\quad w(p)=\hat\var(p)\phi\sin(\phi f(p)),\quad\phi=\frac1{h^{1/3}}.\label{W}\ee

We have
\be Dw=\phi^2\hat\var\cos(\phi f)Df+\phi\sin(\phi f)D\hat\var,\label{dw2}\ee
\beq D^2w&&=-\phi^3\hat\var\sin(\phi f)Df\otimes Df+\phi\sin(\phi f)D^2\hat\var\nonumber\\
&&\quad+\phi^2[\cos(\phi f)Df\otimes D\hat\var+\cos(\phi f)D\hat\var\otimes Df],\label{w1}\eeq and
\beq DW&&=-\frac{w}{v}Z\otimes Df+\frac{\cos(\phi f)}{v}Z\otimes D\hat\var -\frac{\hat\var\cos(\phi f)}{v^2}Z\otimes Dv-tD^2w.\label{DW}\eeq It follows from (\ref{DW}) and(\ref{Pi5}) that
\beq\Upsilon(y)&&=\frac{\cos(\phi f)}{v}\sym Z\otimes D\hat\var -\frac{\hat\var\cos(\phi f)}{v^2}\sym Z\otimes Dv-tD^2w,\label{DW1}\eeq where $\Upsilon(y)$ is given in (\ref{2.7}). In addition, from (\ref{dw2}) and (\ref{w1}),  we have
\be h^{2/3}|Dw|\leq C,\quad |t||D^2w|\leq C\qfq p\in \overline{S},\quad |t|\leq h/2,\quad h>0.\label{dw}\ee

Let $\psi(p)=x$ be the asymptotic coordinate system with $\psi(p_0)=0,$ given in (\ref{o}). Let $\delta_0>0$ be given small such that
$$(0,\delta_0)^2\subset \psi(\B(p_0,\delta)).$$
From (\ref{w1}) and (\ref{pl}), we obtain
\beq C\geq h^2\|D^2w\|^2_{L^2(S)}&&\geq h^2\int_{\B(p_0,\delta)}|D^2w|^2dg=\int_{\B(p_0,\delta)}\sin^2(\phi f)|Df|^2dg\nonumber\\
&&\geq\si\int_{(0,\delta_0)^2}\sin^2(\phi x_1)dx_1dx_2=\si\delta_0\int_0^{\delta_0}\sin^2(\phi x_1)dx_1\nonumber\\
&&=\frac{\si\delta_0}2\int_0^{\delta_0}[1-\cos(2\phi x_1)]dx_1\nonumber\\
&&\geq\si\sum_{k=1}^m\int_{h^{1/3}(-\frac14+k)\pi}^{h^{1/3}(\frac14+k)\pi}dx_1=\frac{\si h^{1/3}\pi}2m\nonumber\\
&&\geq\si(\delta_0-\frac{5h^{1/3}\pi}4)\geq\si\qfq h>0\quad\mbox{small},\label{D^2w}\eeq
where
$$m=\Big[\frac{\delta_0}{h^{1/3}\pi}-\frac14\Big].$$ A similar argument as above yields
\be h^{4/3}\|Dw\|^2_{L^2(S)}\geq\int_{\B(p_0,\delta)}\cos^2(\phi f)|Df|^2dg\geq\si\qfq h>0\quad\mbox{small}\label{dw1}. \ee

 Noting that $|t|\leq h/2,$  from (\ref{DW1}) and (\ref{D^2w}), we have
\be \|\Upsilon(y)\|^2_{L^2(S)}\leq C\qfq h>0\quad\mbox{small}.\label{Upsilon}\ee

Using the formulas (\ref{2.5}), (\ref{DW}), and (\ref{dw}), we obtain
\beq(1-Ch^2)|\nabla y|^2&&\leq C(|DW|^2+h^2|D^2w|^2+|w|^2+|Dw|^2+|Z|^2)\nonumber\\
&&\leq \frac{C}{h^{4/3}}\qfq z=p+t\n\in\Om,\quad|t|\leq h/2,\quad h>0,\label{nablay}\eeq which gives
$$
\|\nabla y\|^2_{L^2(S)}\leq \frac{C}{h^{4/3}}.$$
Moreover, it follows from (\ref{2.5}) and (\ref{dw1}) that
$$\|\nabla y\|^2_{L^2(S)}\geq\si\|\nabla y+tP(y)\|^2_{L^2(S)}\geq\si\|Dw\|^2_{L^2(S)}\geq\frac{\si}{h^{4/3}},$$ that is,
\be \frac{\si}{h^{4/3}}\leq\|\nabla y\|^2_{L^2(S)}\leq\frac C{h^{4/3}}\qfq h>0\quad\mbox{small},\quad|t|\leq h/2.\label{delta y}\ee

Let $X(y)$ be given in (\ref{2.7}).
It follows from (\ref{W}) and (\ref{dw}) that
\beq\|X(y)\|^2_{L^2(S)}&&=\|\ii(W)\Pi\|^2_{L^2(S)}=\|\frac{\hat\var\cos(\phi f)}{v}\ii(Z)\Pi-\ii(tDw)\Pi\|^2_{L^2(S)}\leq C.\label{Upsilon1}\eeq In addition, by an argument as for (\ref{D^2w}), we have
\beq\|X(y)\|^2_{L^2(S)}\geq&&\si\int_{\B(p_0,\delta)}\cos^2(\phi f)||[(\nabla\n)^2]^{1/2}Df|^2dg-Ch^{1/3}\nonumber\\
&&\geq\si-Ch^{1/3}\geq\si\qfq h>0\quad\mbox{small}.\label{Upsilon2}\eeq
From (\ref{2.6}), (\ref{Upsilon}), (\ref{delta y}), and (\ref{Upsilon1}), we obtain
\beq\|\sym\nabla y\|^2_{L^2(S)}&&\leq 2\|\sym\nabla y+t\sym P(y)\|^2_{L^2(S)}+Ch^2\|\nabla y\|^2_{L^2(S)}\leq C,\label{u1}\eeq
and, by (\ref{delta y})  and (\ref{Upsilon2}),
\beq 2\|\sym\nabla y\|^2_{L^2(S)}&&\geq\|\sym\nabla y+t\sym P(y)\|^2_{L^2(S)}-Ch^2\|\nabla y\|^2_{L^2(S)}\nonumber\\
&&\geq\frac12\|X(y)\|_{L^2(S)}^2-Ch^{2/3}\geq\si,\label{Phi1}\eeq respectively.

Now we consider the  ansatze for the geometry rigidity inequality, given by
\be u(z)=z+h^\tau y(z),\quad \tau>4/3,\ee where $y$ is given in (\ref{u}).  Then
$$\nabla u-I=h^\tau\nabla y.$$ Let $\Phi(B)$ be given in (\ref{Phi}). From (\ref{nablay}) and (\ref{u1}), we have
\beq\|\Phi(h^\tau\nabla y)\|^2_{L^2(S)}&&=h^{2\tau}\int_S|\sym\nabla y+\frac{h^\tau}2\nabla^Ty\nabla y|^2dg\nonumber\\
&&\leq h^{2\tau}(2\|\sym\nabla y\|^2_{L^2(S)}+h^{2\tau}\int_S|\nabla y|^4dg)\nonumber\\
&&\leq  C(1+h^{2(\tau-4/3)})h^{2\tau}\leq Ch^{2\tau},\label{Phi3}\eeq and from (\ref{Phi1})
\be\|\Phi(h^\tau\nabla y)\|^2_{L^2(S)}\geq h^{2\tau}(\si-Ch^{2(\tau-4/3)})\geq\si h^{2\tau},\label{Phi4}\ee respectively.

 We set $A=\nabla u$ and $B=h^\tau\nabla y$ in Lemma \ref{l2.4} and use
(\ref{A2}), (\ref{2.2}), and (\ref{Phi3}) to obtain
\beq \|\dist(\nabla u,\SO(3))\|^2_{L^2(\Om)}&&\leq 2\sqrt3\|\Phi(h^\tau\nabla y)\|^2_{L^2(\Om)}\leq Ch^{1+2\tau}.\label{dist}\eeq Similarly, it follows from (\ref{A2}), (\ref{2.2}), (\ref{nablay}) and (\ref{Phi4}) that
\beq\|\dist(\nabla u,\SO(3))\|^2_{L^2(\Om)}\geq\si\|\Phi(h^\tau\nabla y)\|^2_{L^2(\Om)}\geq\si h^{1+2\tau}.\label{dist1}\eeq

Finally, using (\ref{dist}), (\ref{dist1}), and (\ref{delta y}), we obtain
$$\frac{\si}{h^{4/3}}\leq\frac{\|\nabla u-I\|^2_{L^2(\Om)}}{\|\dist(\nabla u,\SO(3))\|^2_{L^2(\Om)}}
=\frac{h^{2\tau}\|\nabla y\|^2_{L^2(\Om)}}{\|\dist(\nabla u,\SO(3))\|^2_{L^2(\Om)}}\leq\frac C{h^{4/3}}.$$
The estimate (\ref{e2}) follows. \\

(b)\,\,\,Let $S$ be elliptic. We look for the ansatz in the form
\be u(z)=z+h^\tau y(z)\qfq z=p+tn(p)\in\Om,\quad \tau>1,\ee
where
\be y=-tDw+w\n\label{y2}\ee is the ansatz, given in the proof \cite[Theorem 1.4]{Yao20181} for the optimal constant of the Korn inequality. From \cite{Yao20181}, we have
\be\|\nabla y\|_{L^\infty(\Om)}\leq\frac C{h^{1/2}},\quad \frac{\si}{h^{1/2}}\leq\frac{\|\nabla y\|_{L^2(\Om)}}{\|\sym\nabla y\|_{L^2(\Om)}}\leq\frac{C}{h^{1/2}}.\label{ga20}\ee
We have
$$\nabla u-I=h^\tau\nabla y,\quad \Phi(\nabla u-I)=h^\tau(\sym\nabla y+\frac{h^\tau}2\nabla^Ty\nabla y).$$
It follows from (\ref{ga20}) that
$$ \frac{\si}{h^{1/2}(1+h^{\tau-1})}\leq\frac{\|\nabla u-I\|_{L^2(\Om)}}{\|\Phi(\nabla u-I)\|_{L^2(\Om)}}\leq \frac{C}{h^{1/2}(1-Ch^{\tau-1})}. $$
Thus the proof is complete by Lemma \ref{l2.4}.
$\Box$\\

{\bf Proof of Proposition \ref{p1}}\,\,\,Let $p\in M$ be given and let $e\in T_{p}M$ be such that
$$\nabla_e\n=0,\quad |e|=1.$$
Consider the geodesic
\be\gamma(t,p)=\exp_{p}te\qfq t\in R\label{ga}\ee on $M$ in the induced metric $g,$  where $\exp_q:$ $T_pM\rw M$ is the exponential map. We will show that $\g(t,p)$ satisfies (\ref{pra}).

From \cite[Lemma 2.7]{Yao20181}, there are a neighbourhood $N$ of $p$ and a vector field $X$ such that
\be X(p)=e,\quad |X|=1,\quad \nabla_X\n=0\qfq q\in N.\label{ga8}\ee Set
$$Y=QX$$ where¡¡ $Q$ is given in (\ref{1.3n}). Then
$$\nabla_Y\n=\lam Y,\quad\<X,Y\>=0,\quad|Y|=1\qfq q\in N,$$ where $\lam\not=0$ is the nonzero principal curvature.

We have
\beq\nabla_X\nabla_Y\n&&=\nabla_X(\lam Y)=X(\lam)Y+\lam \nabla_XY=X(\lam)Y+\lam\<\nabla_XY,X\>X+\lam\<\nabla_XY,\n\>\n\nonumber\\
&&=X(\lam)Y+\lam\<D_XY,X\>X.\nonumber\eeq Thus
$$X(\lam)Y+\lam\<D_XY,X\>X=\nabla_Y\nabla_X\n+\nabla_{[X,Y]}\n=\lam\<[X,Y],Y\>Y,$$ which yields, by $\lam\not=0,$
$$X(\lam)=\lam\<[X,Y],Y\>,\quad\<D_XY,X\>=0\qfq q\in N.$$
We obtain
\be D_XX=\<D_XX,Y\>Y=-\<X,D_XY\>Y=0\qfq q\in N,\label{ga1}\ee
\be D_XY=\<D_XY,X\>X=0\qfq q\in N.\ee

Consider the flow by $X:$
$$\a'(t,q)=X(\a(t,q)),\quad \a(0,q)=q\qfq q\in N.$$  Set
\be\g(t,q)=\exp_qtX(q)\qfq q\in N.\label{ga5}\ee The formula (\ref{ga1}) shows that
\be\g(t,q)=\a(t,q),\quad\dot\g(t,q)=X(\g(t,q))\quad\mbox{when}\quad \a(t,q)\in N.\label{ga2}\ee

Next, we prove that
\be \nabla_{\dot\g(t,q)}\n=0\qfq (t,q)\in\R\times N.\label{ga3}\ee
Let $q\in N$ be given. From (\ref{ga2}) and (\ref{ga8}), there is a largest number $\epsilon_0>0$ such that (\ref{ga3}) hold for all $t\in[0,\varepsilon_0).$ Let $\varepsilon_0<\infty.$ Clearly
$$\nabla_{\dot\g(\varepsilon_0,q)}\n=0.$$
By \cite[Lemma 2.7]{Yao20181} again, there is a vector field $Z$ on  a neighbourhood of $\g(\varepsilon_0,q)$ such that
$$Z(\g(\varepsilon_0,q))=\dot\g(\varepsilon_0,q),\quad |Z|=1,\quad \nabla_Z\n=0.$$
From the uniqueness of a geodesic, (\ref{ga3}) would hold true for all $t\in[0,\varepsilon_1)$ for some $\varepsilon_1>\varepsilon_0.$ This contradiction shows that (\ref{ga3}) hold for all $t\in[0,\infty).$ A similar argument shows
that (\ref{ga3}) also hold for all $t\in(-\infty,0].$

We extend the vector field $X$ from $N$ to $\hat N,$ still denoted by $X,$ by
$$X(\g(t,q))=\dot\g(t,q)\qfq (t,q)\in\R\times N,$$ where
$$\hat N=\{\,\g(t,q)\,|\,t\in\R,\,\,q\in N\,\}.$$ Moreover, we extend $Y$ from $N$ to $\hat N$ by parallelling translation $Y(q)$ to $Y(\g(t,q))$ along the geodesic $\g(t,q)$ in the induced metric of $M.$ Then $X$ and $Y$ forms an orthonormal frame on
$\hat N$ with
$$\nabla_X\n=0,\quad\nabla_Y\n=\lam Y,\quad\lam\not=0,\quad D_XX=D_XY=0\qfq q\in\hat N.$$

Let
$$\g(t,q)=\g_1(t)X+\g_2(t)Y+\g_3(t)\n.$$ Then
\beq\dot\g(t,q)&&=\g_1'(t)X+\g_2'(t)Y+\g_3'(t)\n+\g_1(t)\nabla_{X}X+\g_2(t)\nabla_XY+\g_3(t)\nabla_X\n\nonumber\\
&&=\g_1'(t)X+\g_2'(t)Y+\g_3'(t)\n\qfq t\in\R.\nonumber\eeq
Thus the formulas $\g'(t,q)=X(\g'(t,q))$ imply that
$$      \g_1(t)=\g_1(0)+t,\quad\g_2(t)=\g_2(0),\quad\g_3(t)=\g_3(0)\qfq t\in\R.             $$
Thus $\g(t,q)$ is a straight line in $\R^3$  for given $q\in N.$ The proof is complete. \hfill$\Box$

\begin{lem}\label{lg2.5} Let $p_0\in S$ be such that $(\ref{ga7})$ holds.  For any $a>0,$ there exists a principal coordinate system $\psi^{-1}:$  $\in(-a,a)\times(-\varepsilon,\varepsilon)\rw M$ such that
\be \psi(\g(t,p_0))=(t,0)\qfq t\in(-a,a)\label{ga6}\ee where $\varepsilon>0$ is a number small.
\end{lem}

{\bf Proof}\,\,\,Let $\g(t,q),$ $X,$ and $Y$ be given in the proof of Proposition \ref{p1}.
Define
\be\eta(\g(t,q))=e^{\int_0^t\<D_YX,Y\>\circ\g(s,q)ds}\qfq \g(t,q)\in\hat N.\label{ga10}\ee Then
\be[X,\eta Y]=D_X(\eta Y)-\eta D_YX=[X(\eta)-\eta\<D_YX,Y\>]Y=0\qfq q\in\hat N.\label{ga4}\ee
Let $\b(s,q)$ be the flow by the vector $\eta Y,$ i.e., for each $q\in\hat N,$ there is $\varepsilon(q)>0$ such that
$$\dot\b(s,q)=\eta Y(\b(s,q)),\quad \b(0,q)=q\qfq s\in(-\varepsilon(q),\varepsilon(q)).$$ Since the interval $[-a,a]$ is compact,
there is a constant $\varepsilon>0$ small such that
$$\dot\b(s,\g(t,p_0))=\eta Y(\b(s,\g(t,p_0)))\quad\mbox{for all}\quad (t,s)\in(-a,a)\times(-\varepsilon,\varepsilon).$$
From \cite[p.233, Theorem 9.44]{Lee}, (\ref{ga4}) implies that
\be\g(t,\b(s,p_0))=\b(s,\g(t,p_0))\quad\mbox{for all}\quad (t,s)\in(-a,a)\times(-\varepsilon,\varepsilon).\label{ga9}\ee

We define $\psi^{-1}:$ $(-a,a)\times(-\varepsilon,\varepsilon)\rw M$ by
$$\psi^{-1}(x_1,x_2)=\g(x_1,\b(x_2,p_0)).$$
From Proposition \ref{p1}, there is a $\varepsilon>0$ such that $\psi(q)=x$ defines a coordinate satisfying  (\ref{ga6}). Furthermore,
(\ref{ga9}) implies that $\pl x_1=X$ and $\pl x_2=\eta Y.$  \hfill$\Box$\\

We make some further preparations for the proof of Theorem \ref{t1.2}.
Let assumption (\ref{ga7}) hold. Let $a>\max\{|t_-|,\,|t_+|\}$ be given. Let $\psi(q)=(x_1,x_2)$ be the principal coordinate given in Lemma \ref{lg2.5}. Let $X$ and $\eta Y$ be the vector field given in Lemma \ref{lg2.5} such that
\be\pl x_1=X,\quad \pl x_2=\eta Y.\label{ga12}\ee Then
\be D_XX=D_XY=0,\quad D_YX=\varrho Y,\quad D_YY=-\varrho X,\label{ga11}\ee where $\varrho=\<D_YX,Y\>.$
For $x_2\in(-\varepsilon,\varepsilon),$ let $t_+(x_2)>0>t_-(x_2)$ be such that
$$\g(t_\pm(x_2),\b(x_2,p_0))\in\pl S,\quad \g(x_1,\b(s,p_0))\in S\qfq x_1\in(t_-(x_2),t_+(x_2)).$$ Set
$$S_0=\{\,\g(x_1,\b(x_2,p_0))\,|\,(x_1,x_2)\in(t_-(x_2),t_+(x_2))\times(-\varepsilon,\varepsilon)\,\}.$$
Then $S_0\subset S.$

We will construct the ansatz with its values supported on $S_0.$ Let the functions $\eta$ and $\varrho$ be given in (\ref{ga10}) and (\ref{ga11}), respectively, on $S_0.$ Set
$$\varpi(p)=e^{\int_0^{x_1}\varrho(s,x_2)ds}\qfq p\in S_0.$$ We further define
\be v(p)=\varpi\var'(x_2),\quad b(p)=-\frac1\lam Y(v),\quad w(p)=\lam v(p)-Y(b)\qfq p\in S_0. \label{ga15}\ee Then
\be v_{x_1}=v\varrho,\quad Y(v)+b\lam=0,\quad Y(b)-\lam v+w=0\qfq p\in S_0. \label{ga13}\ee

Consider the ansatz
\be y(z)=\left\{\begin{array}{l}V(z)+tW(z)+b\n\qfq z=p+t\n,\quad p\in S_0,\,\,|t|<h/2,\\
0,\quad z=p+t\n,\quad p\in S/S_0,\,\,|t|<h/2,\end{array}\right.\label{ga14}\ee
where
$$ V=vY,\quad W=-b_{x_1}X+wY\quad p\in S_0.$$
It follows from (\ref{ga11}) and (\ref{ga12}) that
$$D_XV=v_{x_1}Y,\quad D_YV=-v\varrho X+Y(v)Y,$$
$$D_XW=-b_{x_1x_1}X+w_{x_1}Y,\quad D_YW=-[Y(b_{x_1})+\varrho w]X+[Y(w)-b_{x_1}\varrho]Y. $$
Using the above formulas and from Lemma \ref{lem2.1}, we obtain
\beq|\nabla y+tP(y)|^2&&=|DV+tDW+b\Pi|^2+|Db-\ii(V+tW)\Pi|^2+|W|^2\nonumber\\
&&=(\<D_XV,X\>+t\<D_XW,X\>)^2+(\<D_XV,Y\>+t\<D_XW,Y\>)^2\nonumber\\
&&\quad+(\<D_YV,X\>+t\<D_YW,X\>)^2+\{Y(v)+t[Y(w)-b_{x_1}\varrho]+b\lam\}^2\nonumber\\
&&\quad+b_{x_1}^2+[Y(b)-(v+tw)\lam]^2+b_{x_1}^2+w^2\nonumber\\
&&=t^2b_{x_1x_1}^2+(v_{x_1}+tw_{x_1})^2+\{v\varrho+t[Y(b_{x_1})+\varrho w]\}^2\nonumber\\
&&\quad+\{Y(v)+t[Y(w)-b_{x_1}\varrho]+b\lam\}^2+2b_{x_1}^2\nonumber\\
&&\quad+[Y(b)-(v+tw)\lam]^2+w^2\qfq p\in S_0,\label{ga16}\eeq where the following formulas have been used
$$\nabla_X\n=0,\quad\nabla_Y\n=\lam Y\qfq p\in S_0.$$
Noting (\ref{ga13}), by a similar computation, we have
\be|\Upsilon(y)|^2=t^2\{b_{x_1x_1}^2+\frac12[w_{x_1}-Y(b_{x_1})-\varrho w£©]^2+[Y(w)-b_{x_1}\varrho]^2\}\qfq p\in S_0,\ee
\be|Db-\ii(V+tW)\Pi+W|^2=[Y(b)-(v+tw)\lam+w]^2=t^2\lam^2w^2\qfq p\in S_0.\label{ga17}\ee

{\bf Proof of Theorem \ref{t1.2}}\,\,\,Let $\var_0\in\CC^4_0(-\varepsilon,\varepsilon)$ be given such that
$$\var_0(x_2)=1\qfq x_2\in(-\varepsilon/2,\varepsilon/2).$$ Set
$$\var(x_2)=\var_0(x_2)\cos(\phi x_2),\quad \phi=\frac1{h^{1/4}},\qfq x_2\in(-\varepsilon,\varepsilon),$$ in (\ref{ga15}).
$$ v(p)=\varpi\var'(x_2),\quad b(p)=-\frac1\lam Y(v),\quad w(p)=\lam v(p)-Y(b),$$Then
it follows from (\ref{ga15}) that
$$v(p)=-\frac{\varpi\var_0(x_2)\sin(\phi x_2)}{h^{1/4}}+\o(\frac1{h^{1/4}}),\quad  b(p)=\frac{\varpi\var_0(x_2)\cos(\phi x_2)}{\lam\eta h^{1/2}}+\o(\frac1{h^{1/2}}),$$
$$ w(p)=\frac{\varpi\var_0(x_2)\cos(\phi x_2)}{\lam\eta^2 h^{3/4}}+\o(\frac1{h^{3/4}}),\quad Y(w)=-\frac{\varpi\var_0(x_2)\sin(\phi x_2)}{\lam\eta^3 h}+\o(\frac1{h}).$$
Noting $|t|\leq h/2$ and from (\ref{ga16})-(\ref{ga17}) and (\ref{2.6}), we obtain
$$\frac{\si\var_0^2(x_2)\cos^2(\phi x_2)}{ h^{3/2}}\leq|\nabla y+tP(y)|^2\leq\frac{C}{h^{3/2}}\qfq p\in S_0,$$
$$\frac{\si t^2\var_0^2(x_2)\sin^2(\phi x_2)}{h^2}\leq|\sym\nabla y+t\sym P(y)|^2\leq \frac{Ct^2}{h^2}\qfq p\in S_0.$$
Moreover, a similar argument as in (\ref{D^2w}) yields
$$\int_{S_0}\var_0^2(x_2)\sin^2(\phi x_2)dg,\quad\int_{S_0}\var_0^2(x_2)\cos^2(\phi x_2)dg\geq\si\varepsilon-\si_0h^{1/4}.$$
Using the above formulas, we have
\be|\nabla y|^2\leq \frac C{h^{3/2}}\qfq z=p+t\n\in\Om,\label{ga18}\ee
\be\frac{\si }{h^{3/2}}\leq\|\nabla y\|^2_{L^2(S)}\leq \frac C{h^{3/2}},\quad\frac{\si t^2}{h^2}\leq \|\sym\nabla y\|^2_{L^2(S)}\leq\frac{Ct^2}{h^2}. \label{ga19}  \ee

Consider the ansatz
$$u(z)=z+h^\tau y(z),\quad \tau>\frac32,\qfq z=p+t\n\in\Om,$$ where $y$ is given in (\ref{ga14}). Then
$$\nabla u-I=h^\tau\nabla y,\quad\Phi(\nabla u-I)=h^\tau(\sym\nabla y+\frac{h^\tau}2\nabla^Ty\nabla y).$$
It follows from (\ref{ga18}) and (\ref{ga19}) that
$$\si h^{2\tau-\frac12}\leq\|\nabla u-I\|^2_{L^2(\Om)}\leq Ch^{2\tau-\frac12},$$
$$\si h^{2\tau}(h-h^{2\tau-2})\leq\|\Phi(\nabla u-I)\|_{L^2(\Om)}^2\leq Ch^{2\tau}(h+h^{2\tau-2}).$$
Since $\det\nabla u=\det(I+h^\tau\nabla y)>0$ when $h>0$ is small enough, from Lemma \ref{l2.4} and the above estimates, we obtain
$$\frac{\si}{h^{3/2}(1+h^{2(\tau-3/2)})}\leq\frac{\|\nabla u-I\|^2_{L^2(\Om)}}{\|\dist(\nabla u,\SO(3))\|^2_{L^2(\Om)}}\leq
\frac{C}{h^{3/2}(1-h^{2(\tau-3/2)})}.$$
The proof is complete. \hfill$\Box$\\

{\bf Conflict of interest statement}

There is no conflict of interests.

Ethical approval: This article does not contain any studies with human participants or animals performed by the author.

 \end{document}